\documentclass[10pt, conference, letterpaper]{IEEEtran}
\usepackage[pdftex]{graphicx}
\usepackage{epstopdf}
\usepackage[cmex10]{amsmath}
\usepackage{cases,enumerate}
\usepackage{subeqnarray,subfigure}
\usepackage{cite,setspace}
\usepackage{amsmath,bm,amssymb,amsthm}
\usepackage{amsfonts}
\usepackage{epsfig}
\usepackage{multirow}
\usepackage{amsmath,color}
\usepackage[linesnumbered,ruled,vlined]{algorithm2e}

\begin{document}
\title{Generative AI-Aided QoE Maximization for RIS-Assisted Digital Twin Interaction
	
}

\author{Jiayuan Chen$^{\ast}$, Yuxiang Li$^{\ast}$, Changyan Yi$^{\ast}$ and Shimin Gong$^{\dagger}$\\
	\IEEEauthorblockA{\text{\small $^{\ast}$College of Computer Science and Technology, Nanjing University of Aeronautics and Astronautics, Nanjing, China} \\
		\small $^{\dagger}$School
		of Intelligent Systems Engineering, Sun Yat-sen University, Shenzhen, China 
		\\
		\text{\small Email: \{jiayuan.chen, yuxiangli, changyan.yi\}@nuaa.edu.cn}, gongshm5@mail.sysu.edu.cn\\
	}
}

 \maketitle

\begin{abstract}
In this paper, we investigate a quality of experience (QoE)-aware resource allocation problem for reconfigurable intelligent surface (RIS)-assisted digital twin (DT) interaction with uncertain evolution. 
In the considered system, mobile users are expected to interact with a DT model maintained on a DT server that is deployed on a base station, via effective uplink and downlink channels assisted by an RIS. 
Our goal is to maximize the sum of all mobile users' joint subjective and objective QoE in DT interactions across various DT scenes, by jointly optimizing phase shift matrix, receive/transmit beamforming matrix, rendering resolution configuration and computing resource allocation. While solving this problem is challenging mainly due to the uncertain evolution of the DT model, which leads to multiple scene-specific problems, and require us to constantly re-solve each of them whenever DT model evolves.
 To this end, leveraging the dynamic optimization capabilities of decision transformers and the generalization strengths of generative artificial intelligence (GAI), we propose a novel GAI-aided approach, called the prompt-guided decision transformer integrated with zero-forcing optimization (PG-ZFO). Simulations are conducted to evaluate the proposed PG-ZFO, demonstrating its effectiveness and superiority over counterparts.  
\end{abstract}

\section{Introduction}

\IEEEPARstart{D}{igital} twin (DT) is a groundbreaking technology that can create high-fidelity and interactive virtual counterparts of physical entities, known as DT models. DT is driving a paradigm shift  across a wide-range of industries, such as remote education, personalized healthcare, and intelligent manufacturing, by delivering DT interaction services powered by these advanced DT models \cite{77}. The native way of providing DT interaction services is through wireless networks, with the reconfigurable intelligent surface (RIS)-assisted DT interactions being particularly important.

Although the RIS-assisted virtual-reality interaction services has been widely studied in literature \cite{86, 87}, they cannot directly suit for the DT interaction services due to their unique and critical features that have not yet been well investigated. Specifically, different from the classical RIS-assisted virtual-reality interaction services, RIS-assisted DT interactions must consider the end-to-end round-trip quality due to their joint uplink and downlink communications. Moreover, beyond the objective experiences of mobile users commonly considered in virtual-reality interaction services, the subjective experiences are vital important in DT interaction services. Additionally, distinct from most of service providers, the DT models evolves uncertainly with the ever-changing corresponding physical entities to real-time map them, resulting in different and uncertainly changing DT scenes in DT interaction services. To fill the gap in the literature, in this paper, we study a quality of experience (QoE)-aware resource allocation problem for RIS-assisted DT interaction with uncertain evolution, with the goal of maximizing the sum of all mobile users' mixed subjective and objective QoE in DT interactions across various DT scenes, by determining the phase shift matrix, receive/transmit beamforming matrix, rendering resolution configuration and computing resource allocation.
However, simultaneously considering all aforementioned  features of RIS-assisted DT interaction in our problem is very challenging because of the following reasons.
i) The decision variables are not only high-dimensional but also tightly coupled, meaning that they have to be jointly optimized.
ii) The uncertain evolution of DT models leads to multiple scene-specific  problems, each of them with distinct problem formulation. This requires us to constantly re-solve each of them whenever the DT model evolves.

To tackle these challenges, in this paper, we propose a novel generative artificial intelligence (GAI)-aided approach to solve QoE-aware resource allocation problem for RIS-assisted DT interaction with uncertain evolution. 
In the considered system, mobile users are expected to interact with a DT model maintained on a DT server that is deployed on a base station, via effective uplink and downlink channels assisted by an RIS. By taking into account the uncertain evolution of the DT model, we formulate a QoE-aware resource allocation problem, which consists of unbounded growing amount of different scene-specific problems. 
We aim to maximize the sum of all mobile users' mixed subjective and objective QoE in DT interactions across all scene-specific problems. To circumvent the difficulty in solving this complicated problem, we propose a GAI-aided approach, called prompt-guided decision transformer integrated with zero-forcing optimization (PG-ZFO). Specifically, we first reformulate each individual scene-specific problem into an Markov decision process (MDP). Then, we design a prompt to elaborately capture the scene-specific information. Based on this, we build a prompt-guided decision transformer to generate phase shift matrix, rendering resolution configuration and computing resource allocation. On top of this, we further develop a zero-forcing (ZF)-based optimization algorithm to derive receive/transmit beamforming matrix. Such a decision-making process is iteratively conducted during the DT scene, and thus the scene-specific problem is solved. 
The most significant advantage of the proposed PG-ZFO is its superior generalization ability, which enables it to efficiently solve various scene-specific problems without the need for time-consuming re-training.

The main contributions of this paper are summarized in the following. i) To the best of our knowledge, we are the first to study the RIS-assisted DT interaction with uncertain evolution. We formulate a QoE-aware resource allocation problem, consisting of unbounded growing amount of different scene-specific problems caused by the uncertain evolution of the DT model.  
ii) We propose a novel GAI-aided approach, called PG-ZFO. With its strong generalization capabilities, PG-ZFO  efficiently solve various  scene-specific problem triggered by the uncertain evolution of the DT model without time-consuming re-training.
iii) We conduct  simulations to evaluate the performance of proposed PG-ZFO, and the  results have shown its effectiveness and superiority over counterparts.

\section{System Model and Problem Formulation}\label{SSMPF}

\subsection{System Overview}

\begin{figure}[!t]
	\centering
	\includegraphics[width= 0.83\columnwidth]{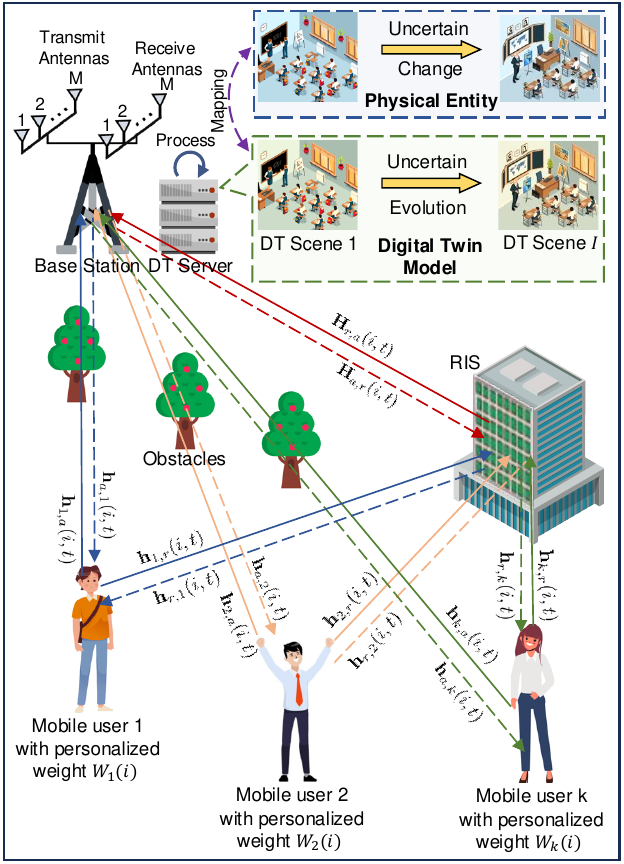} \\
	\caption{An illustration of the considered RIS-assisted DT interaction system.}\label{model1}
\end{figure}

Consider an RIS-assisted DT interaction system, as illustrated in Fig. \ref{model1}, consisting of a DT server $a$ deployed on a base station with a uniform linear array (ULA) of $M$ receive/transmit antennas, located at $\mathbf{q_a} = [x_a, y_a, z_a]$, a set of single-antenna mobile users $\mathcal{K}$ with $|\mathcal{K}| = K$, and an RIS $r$ with $N$ passive reflecting elements at $\mathbf{q_r} = [x_r, y_r, z_r]$. 
The DT server maintains a high-fidelity and interactive DT model
that mirrors the real-time status of a physical entity to provide engaging DT interaction services \cite{1}. To ensure seamless connectivity and real-time interaction, the RIS assists uplink and downlink.

It is worth noting that the DT model uncertainly evolves with the ever-changing physical entity. Consequently, the DT scene $i \in \mathcal{I}$ in which users engage continuously evolves, with the cardinality of $\mathcal{I}$ grows with the evolution, i.e., $|\mathcal{I}|\to+\infty$. Mobile users interact with the DT server at a frequency $\mathcal{F}_i$ during each DT scene $i \in \mathcal{I}$. dividing it into $T_i=1/ \mathcal{F}_i $ time slots with $\mathcal{T}_i=\{1,2,...,T_i\}$. Moreover, each mobile user $k \in \mathcal{K}$ has a personalized weight for subjective and objective experiences, defined as $\mathcal{W}_k(i) = (\varpi_{k}^{\epsilon}(i), \varpi_{k}^{\iota}(i))$, where $\varpi_{k}^{\epsilon}(i)$ and $\varpi_{k}^{\iota}(i)$ represent their degrees of attentions to subjective and objective experiences towards DT scene $i \in \mathcal{I}$, respectively, and satisfying $\varpi_{k}^{\epsilon}(i)+\varpi_{k}^{\iota}(i)=1$.

\subsection{Uplink and Downlink Transmissions in DT Interaction}

Within any time slot $t\in\mathcal{T}_i$ when engaging in DT scene $i\in\mathcal{I}$, each mobile user $k \in \mathcal{K}$, located at $\mathbf{q_{k}}(i) = [x_{k}(i), y_{k}(i), z_{k}(i)]$ transmits its interaction signals to the DT server through an effective uplink channel $\mathbf{h}_{k, a}^{UL}(i, t) \in \mathbb{C}^{M \times 1}$, while obtaining feedback signals from the DT server through an effective downlink channel $\mathbf{h}_{a, k}^{DL}(i, t) \in \mathbb{C}^{M \times 1}$, both assisted by RIS, which can be respectively expressed as
$\mathbf{h}_{k, a}^{UL}(i, t) = \mathbf{h}_{k, a}(i, t)+\mathbf{H}_{r, a}(i, t) \boldsymbol{\Theta}(i, t)\mathbf{h}_{k, r}(i, t)$, and
$
	\mathbf{h}_{a, k}^{DL}(i, t) = \mathbf{h}_{a, k}(i, t)+\mathbf{H}_{a, r}(i, t)\boldsymbol{\Theta}(i, t)\mathbf{h}_{r, k}(i, t), $
where vectors $\mathbf{h}_{k, a}(i, t)$ and $\mathbf{h}_{a, k}(i, t)$ specify the direct uplink and downlink channels between mobile user $k \in \mathcal{K}$ and the DT server; matrices $\mathbf{H}_{r, a}(i, t)$, $\mathbf{H}_{a, r}(i, t)$ indicate the uplink and downlink channels between the RIS and the DT server; vectors $\mathbf{h}_{k, r}(i, t)$, $\mathbf{h}_{r, k}(i, t)$ stand for the uplink and downlink channels between mobile user $k$ and the RIS; $\boldsymbol{\Theta}(i, t) \triangleq diag(A_1(i, t)e^{j\theta_1(i, t)}, ..., A_N(i, t)e^{j\theta_N(i, t)}) \in \mathbb{C}^{N \times N}$ is the phase shift matrix of the RIS, where $A_n(i, t)$ and $\theta_n(i, t)$ represent the amplitude reflection coefficient and the phase shift induced by $n$-th RIS element. Similar to \cite{3, 10}, ideal reflections are assumed, satisfying $|A_n(i, t)e^{j\theta_n(i, t)}|^2 = 1$.

For direct uplink communications of the DT interaction, the channel between mobile user $k \in \mathcal{K}$ and the DT server is modeled as $\mathbf{h}_{k, a}(i, t)=\sqrt{\rho d_{k, a}^{-\alpha_{k, a}}(i)}\mathbf{h}_{k, a}^{NLoS}(i, t)$, where $\rho$ is the path loss coefficient, $\alpha_{k, a}$ is the path loss exponent, $d_{k, a}(i) = \sqrt{||\mathbf{q_{k}}(i) - \mathbf{q_a}||^{2}}$ is the distance between mobile user $k$ and the DT server, and $\mathbf{h}_{k, a}^{NLoS}(i, t) \sim \mathcal{CN}(0, 1)$ is the non-LoS (NLoS) component \cite{26}. 
Meanwhile, the channel between mobile user $k \in \mathcal{K}$ and the RIS is given by $\mathbf{h}_{k, r}(i, t)=\sqrt{\rho d_{k, r}^{-\alpha_{k, r}}(i)}(\sqrt{\frac{G_{k, r}}{G_{k, r}+1}}\mathbf{h}_{k, r}^{LoS}(i)+
	\sqrt{\frac{1}{G_{k, r}+1}}\mathbf{h}_{k, r}^{NLoS}(i, t))$,
where $d_{k, r}(i) = \sqrt{||\mathbf{q_{k}}(i) - \mathbf{q_r}||^{2}}$ is the distance between mobile user $k$ and the RIS, $\mathbf{h}_{k, r}^{NLoS}(i, t) 
 \sim \mathcal{CN}(0, 1)$, and $G_{k, r}$ is the Rician factor. Besides, $\mathbf{h}_{k, r}^{LoS}(i)$ is the LoS component based on  angle of arrival (AoA).
Furthermore, the channel between the RIS and the DT server can be modeled as $\mathbf{H}_{r,a}(i, t)=\sqrt{\rho d_{r,a}^{-\alpha_{r, a}}}(\sqrt{\frac{G_{r, a}}{G_{r, a}+1}}\mathbf{H}_{r, a}^{LoS} 
			+\sqrt{\frac{1}{G_{r, a}+1}}\mathbf{H}_{r, a}^{NLoS}(i, t))$
where $d_{r, a} = \sqrt{||\mathbf{q_{r}} - \mathbf{q_a}||^{2}}$ is the distance between the RIS and the DT server,  $\mathbf{H}_{r, a}^{NLoS}(i, t) \sim \mathcal{CN}(0, 1)$, $G_{r, a}$ is the Rician factor, and $\mathbf{H}_{r, a}^{LoS}$ includes AoA and angle of departure (AoD). 

 Then, we can express the interaction signals of mobile user $k \in \mathcal{K}$ received by the DT server via $\mathbf{h}_{k, a}^{UL}(i, t)$ as $y_{k}^{UL}(i, t)   =  \sqrt{p^{UL}_{k}}(\mathbf{v}_{k}(i, t))^{H}\mathbf{h}_{k, a}^{UL}(i, t)\delta_{k}^{UL}(i, t) + (\mathbf{v}_{k}(i, t))^{H} (\sum\nolimits_{m =1, m \neq k  }^{\mathcal{K}} \sqrt{p_{m}} \mathbf{h}_{m, a}^{UL}(i, t)\delta_{m}^{UL}(i, t))+(\mathbf{v}_{k}(i, t))^{H}(\mathbf{n}_{k}^{UL}(i, t))$, where $p^{UL}_{k}$ is the uplink transmission power, $\delta_{k}^{UL}(i, t) \sim \mathcal{CN}(0, 1)$ is the symbol, $\mathbf{v}_{k}(i, t) \in  \mathbb{C}^{M \times 1}$ is the receive beamforming vector, and $\mathbf{n}_{k}^{UL}(i, t)  \sim \mathcal{CN}(0, (\sigma_{k}^{UL})^2)$ is noise. 

Accordingly, the transmission rate of uplink interaction is  $\upsilon_{k}^{UL}(i, t)=b \log_2(1+\gamma_{k}^{UL}(i, t))$, where $b$ is the uplink channel bandwidth, and $\gamma_{k}^{UL}(i, t)$ is the uplink signal-to-interference-plus-noise ratio (SINR), calculated as $\gamma_{k}^{UL}(i, t)=(p^{UL}_{k}|(\mathbf{h}_{k, a}^{UL}(i, t))^H\mathbf{v}_{k}(i, t)|^2)/((\mathbf{v}_{k}(i, t))^{H}((\sigma_{k}^{UL})^2\mathbf{I}_{M}+ \sum\limits_{m =1, m \neq k  }^{\mathcal{K}} p^{UL}_{m} \mathbf{h}_{m, a}^{UL}(i, t)  (\mathbf{h}_{m, a}^{UL}(i, t))^H)\mathbf{v}_{k}(i, t))$.

Similar to effective uplink channel, we can obtain the downlink transmission rate  as $\upsilon_{k}^{DL}(i, t)=b \log_2(1+\gamma_{k}^{DL}(i, t))$, 
where $\gamma_{k}^{DL}(i, t)=(|(\mathbf{h}_{a, k}^{DL}(i, t))^H\mathbf{w}_{k}(i, t)|^2)/((\sigma_{k}^{DL})^2 + \sum\nolimits_{m =1, m \neq k }^{\mathcal{K}} |(\mathbf{h}_{a, k}^{DL}(i, t))^H\mathbf{w}_{m}(i, t)|^2)$.

\subsection{QoE Model Analysis for DT Interaction} \label{SQoE}
We introduce a novel user-centric QoE metric, i.e., $QoE(\mathcal{W}_{k}(i), t), \forall k \in \mathcal{K}$. Particularly, we consider two key performance indicators, i.e., the Weber-Fechner Law based human perception quality $\mathcal{E}_{k}(i, t)$ and the interaction round-trip latency $\mathcal{L}_{k}(i, t)$, to quantify subjective and objective experiences for DT interactions \cite{90}, respectively. 

According to Weber-Fechner Law \cite{7}, we model the physical stimulus intensity as the rendering resolution $E_{k}(i, t)$, which determines the authenticity of the service. Based on this, $\mathcal{E}_{k}(i, t)=\ln (\frac{E_{k}(i, t)}{E_{min}})$,
where $E_{min}$ is the minimum resolution that mobile users demand.

Within each time slot $t \in \mathcal{T}_i$ when engaging in any DT scene $i \in \mathcal{I}$, each mobile user $k \in \mathcal{K}$ first transmits interaction signal with size $D_{k}^{UL}(i)$  via  $\mathbf{h}_{k, a}^{UL}(i, t)$ to the DT server, with uplink latency $\mathcal{L}^{UL}_{k}(i, t) = D_{k}^{UL}(i) / \upsilon_{k}^{UL}(i, t)$. 

Then, the DT server generates and renders the feedback signal with resolution $E_{k}(i, t)$. The processing latency is $\mathcal{L}_k^{PRO} (i, t) = \xi c E_{k}(i, t)  / f_{k}(i, t)$,
where $\xi$ is the per-sample data size in bits, $c$ is the CPU cycles per bit, and $f_{k}(i, t)$ is the allocated computation frequency. 

Subsequently, the rendered feedback signal with size $\varsigma E_{k}(i, t)$ is  transmitted back to $k \in \mathcal{K}$ via the effective downlink channel $\mathbf{h}_{k, a}^{DL}(i, t)$, resulting in a downlink latency $\mathcal{L}^{DL}_{k}(i, t) = \varsigma E_{k}(i, t) / \upsilon_{k}^{DL}(i, t)$

To sum up, the interaction round-trip latency is $\mathcal{L}_{k}(i, t)=\frac{D_{k}^{UL}(i)}{\upsilon_{k}^{UL}(i, t)} + \frac{\xi  c E_{k}(i, t) }{f_{k}(i, t)}+\frac{\varsigma E_{k}(i, t)}{\upsilon_{k}^{DL}(i, t)}$.

By integrating subjective and objective experience metrics through the personalized weight $\mathcal{W}_k(i)$, we define  $QoE(\mathcal{W}_k(i), t)=\varpi_{k}^{\epsilon}(i)F^{'}(\mathcal{E}_{k}(i, t))+\varpi_{k}^{\iota}(i)F^{''}(\mathcal{L}_{k}(i, t))$, 
where $F^{'}(\mathcal{E}_{k}(i, t))=\frac{\mathcal{E}_{k}(i, t)}{\mathcal{E}_{max}}, F^{''}(\mathcal{L}_{k}(i, t))=1 - \frac{\mathcal{L}_{k}(i, t)}{\mathcal{L}_{max}}$.
 $\mathcal{E}_{max}$ is the maximal perception quality that each mobile user can achieve, and $\mathcal{L}_{max}$ is the maximal interaction round-trip latency that each mobile user can tolerate.



\subsection{Problem Formulation}\label{ProFor}

For engaging in each DT scene $i \in \mathcal{I}$, we aim to maximize the sum of all mobile users' QoE of DT interaction across all time slots, by jointly optimizing phase shift matrix $\boldsymbol{\Theta}(i, t)$, receive beamforming matrix $\mathbf{V}(i, t):=\{\mathbf{v}_{1}(i, t), ..., \mathbf{v}_{K}(i, t)\}$ and transmit beamforming matrix $\mathbf{W}(i, t):=\{\mathbf{w}_{1}(i, t), ..., \mathbf{w}_{K}(i, t)\}$, rendering resolution configuration $E_{k}(i, t)$ and computing resource allocation $f_{k}(i, t)$, i.e.,
	\begin{align}
	\mathcal{P}(i)\hspace{-0.6mm}:&\max_{\boldsymbol{\Theta}(i, t), \mathbf{V}(i, t), \mathbf{W}(i, t), E_{k}(i, t), f_{k}(i, t)} \hspace{-1mm}\sum_{t\in\mathcal{T}_i}\sum_{k \in \mathcal{K}}{QoE(\mathcal{W}_k(i), t)}  \label{obj} \\
	s.t., ~& \theta_n(i, t)\in[0, 2\pi) ,  \forall n \in \mathcal{N}, i \in \mathcal{I}, t \in \mathcal{T}_i, \tag{\ref{obj}{a}} \label{ris} \\
	& E_{k}(i, t) \in [E_{min}, E_{max}] ,   \forall k \in \mathcal{K}, i \in \mathcal{I}, t \in \mathcal{T}_i, \tag{\ref{obj}{b}} \label{reso} \\
	& \sum_{k \in \mathcal{K}} f_{k}(i, t) \leq C, f_{k}(i, t) > 0, \forall i \in \mathcal{I}, t \in \mathcal{T}_i, \tag{\ref{obj}{c}}\label{comp} \\
	&   \sum_{k \in \mathcal{K}}  || \mathbf{w}_{k}(i, t)||^2 \leq P^{max}, \forall i \in \mathcal{I}, t \in \mathcal{T}_i, \tag{\ref{obj}{d}} \label{tran-ris} \\ 
	& \mathcal{L}_{k}(i, t) \leq \mathcal{L}_{max},  \forall k \in \mathcal{K}, \forall i \in \mathcal{I}, t \in \mathcal{T}_i, \tag{\ref{obj}{e}} \label{laten}
\end{align}
where constraint (\ref{ris}) signifies the continuous phase shift control; (\ref{reso}) defines the range for feedback signal resolution; (\ref{comp}) limits computational resource allocation, with $C$ as the computational capacity of the DT server \cite{89}; (\ref{tran-ris}) imposes a downlink transmission power constraint, with $P^{max}$ as the maximum transmit power of the DT server;  (\ref{laten}) guarantees that the interaction round-trip latency of each mobile user stays within a pre-determined threshold.

It is worth noting that, due to the uncertain DT evolution, the optimization of the considered system actually involves a series of problems different scene-specific problems:
$\boldsymbol{\mathcal{P}} = \{\mathcal{P}(1), \mathcal{P}(2), ..., \mathcal{P}(|\mathcal{I}|)\}_{|\mathcal{I}|\to+\infty}.$ However, solving $\boldsymbol{\mathcal{P}}$ is very challenging because i) 
 each $\mathcal{P}(i) \in \boldsymbol{\mathcal{P}}$ is NP-hard due to  non-convex quadratic terms, and spans multiple time slots requiring a dynamic online algorithm. ii) Constant re-solving $\mathcal{P}(i) \in \boldsymbol{\mathcal{P}}$  is needed as new DT scenes emerge, exacerbating the computational complexity.

For solving different scene-specific problem within $\boldsymbol{\mathcal{P}}$, and inspired by prompt-based learning in GAI models like ChatGPT and DALL-E \cite{16} , we extend the traditional decision transformer \cite{38} to a prompt-guided decision transformer. This model leverages prompts  to encode and exploit scene-specific information, enabling it to generalize and efficiently solve new scene-specific problems without re-training after being trained on historical data.

\section{Prompt-Guided  Decision Transformer Integrated with  Zero-Forcing Optimization} \label{solu}

\subsection{Problem Reformulation}\label{reformul}
We reformulate each scene-specific  problem $\mathcal{P}(i) \in \bm{\mathcal{P}}$ into an MDP $\mathcal{M}(i) \in  \bm{\mathcal{M}}$ over time slot $\mathcal{T}_i, \forall i \in \mathcal{I}$. Each MDP $\mathcal{M}(i) \in  \bm{\mathcal{M}}$ is formulated as follows:

\textbf{States}: Within each time slot $t \in \mathcal{T}_i$, the current state $s(i, t)$ can be represented as $s(i, t)  =  \{ \mathcal{W}_k(i), D^{UL}_k(i), \mathbf{h}_{k, a}^{UL}(i, t), \mathbf{h}_{a, k}^{DL}(i, t), QoE(\mathcal{W}_k(i), t-$$1)\}_{\forall  k \in \mathcal{K}}$.

 \textbf{Actions:} The action space is denoted as $A(i)$. Within time slot $t \in \mathcal{T}_i$, the action $a(i, t) $ can be defined as $a(i, t) =  \{  \boldsymbol{\Theta}(i, t), \{E_k(i, t)\}_{\forall k \in \mathcal{K}}, \{f_{k}(i, t)\}_{\forall k \in \mathcal{K}},  \mathbf{V}(i, t), \mathbf{W}(i, t) \}$.

\textbf{Rewards:} Observing from the formulation of $\mathcal{P}(i) \in 	\boldsymbol{\mathcal{P}}$, within each time slot $t \in \mathcal{T}_i$, reward $r(i, t) = \sum\nolimits_{k \in \mathcal{K}}{QoE(\mathcal{W}_k(i), t)} - \delta 	\sum\nolimits_{k \in \mathcal{K}}{l_k(i, t)}$ if (\ref{tran-ris}) is met, and $r(i, t)=	-\sum\nolimits_{k \in \mathcal{K}}{QoE(\mathcal{W}_k(i), t)}$ otherwise,
	where $-\sum_{k \in \mathcal{K}}{QoE(\mathcal{W}_k(i), t)}$ and $-\delta \sum_{k \in \mathcal{K}}{l_k(i, t)}$ are the penalties for violating constraints (\ref{tran-ris}) and (\ref{laten}), respectively. $\delta $ is the penalty coefficient, and the penalty term $l_k(i, t) = 0$ if $ \mathcal{L}_k(i) \leq \mathcal{L}_{max}$, and $l_k(i, t) = 	\mathcal{L}_{max}$ otherwise.
	
 \textbf{State transition probabilities:} The state transition probabilities is denoted as $Pr(i, s(i, t+1)|s(i, t), a(i, t)) \in [0, 1]$.

\subsection{Prompt-Guided Decision Transformer in PG-ZFO}
\subsubsection{Design of Prompt}
We use ``decision-making trajectory" in designing the prompt. Specifically, the prompt $\tau_{\star}(i)$ consists of a sequence of decision-making trajectory tuples. Following this, $\tau_{\star}(i) =  ( r_{\star}(i, t), s_{\star}(i, t), \vartheta_{\star} (i, t), ...,  r_{\star}(i,  t + T_{\star}), s_{\star}(i, t + T_{\star}), \vartheta_{\star} (i,  t + T_{\star}))$, where $ T_{\star} < T_i$ denotes the amount of  tuples stored in the prompt, and the subscript $\cdot_{\star}$ indicates  the elements related to prompts.  $\tau_{\star}(i)$ implicitly captures the reward function and state transition probabilities of $\mathcal{M}(i) \in \bm{\mathcal{M}}$, thereby encoding the scene-specific information of $\mathcal{P}(i) \in \bm{\mathcal{P}}$. To offer the prompt-guided decision transformer a more forward-looking scene-specific information, we replace $r_{\star}(i, t)$ with the returns-to-go (RTG), $\hat{r}_{\star}(i, t)$,  defined as 
$\hat{r}_{\star}(i, t) = \hat{R}(i) - \sum\nolimits_{j = 1}^{j=t}  r_{\star}(i, j)$, where $\hat{R}(i) = \sum\nolimits_{t \in \mathcal{T}_i } \sum\nolimits_{k \in \mathcal{K}}\varpi_{k}^{\epsilon}(i) + \varpi_{k}^{\iota}(i)$ denotes the maximum total rewards over for $\mathcal{M}(i)$.

\subsubsection{Structure of Prompt-Guided Decision Transformer}

\textbf{\emph{Computing embeddings for input tokens}}:
We take the prompt $\tau_{\star}(i)$ of MDP $\mathcal{M}(i) \in \bm{\mathcal{M}}$ and the most recent decision-making trajectory with length of $L$ prior to the current time slot $t \in \mathcal{T}_i$, i.e., $\tau(i, t)$. To process the RTG, states, and decisions in the prompt, which represent three distinct token modalities, we employ modality-specific trainable linear layers for embedding. A trainable linear layer is also used to add positional embeddings to token embeddings within the same decision-making trajectory tuple in the prompt. Similarly, trainable linear layers are applied to $\tau(i, t)$.

\textbf{\emph{Obtaining decisions using causal transformer}}:
The embedded input tokens are fed into a causal transformer  comprising $\mathcal{G}$-stacked identical decoders. Each decoder processes the tokens through a masked multi-head self-attention module to capture dependencies across tokens. A feedforward layer enhances these representations through position-wise transformations, while layer normalization stabilizes training and facilitates gradient updates. The causal transformer processes tokens sequentially across the $\mathcal{G}$-stacked decoders, ultimately outputting the hidden states. Then, we input them to a trainable linear decision prediction layer to generate the decisions.

\subsection{ZF-based Optimization Algorithm Integrated in PG-ZFO}
The prompt-guided decision transformer struggles to generate receive/transmit beamforming matrix due to the curse of dimensionality. To address this, we propose a low-complexity ZF-based optimization algorithm.

For each time slot $t \in \mathcal{T}_i$ of MDP $\mathcal{M}(i) \in \bm{\mathcal{M}}$, given the decisions $\vartheta(i, t) = \{ \boldsymbol{\Theta}(i, t), \{E_k(i, t)\}_{\forall k \in \mathcal{K}}, \{f_{k}(i, t)\}_{\forall k \in \mathcal{K}}\}$, the optimization of $\kappa(i, t) = \{\mathbf{V}(i, t), \mathbf{W}(i, t)\}$ for MDP $\mathcal{M}(i) \in \bm{\mathcal{M}}$ can be formulated as 
	\begin{align}
	&\min_{\mathbf{V}(i, t), \mathbf{W}(i, t)} \sum_{k \in \mathcal{K}}{\varpi_{k}^{\iota}(i)(\frac{D_{k}^{UL}(i)}{\upsilon_{k}^{UL}(i, t)} + \frac{\varsigma E_{k}(i, t)}{\upsilon_{k}^{DL}(i, t)})},  \label{totalaten} \\
	s.t., ~&   \text{(\ref{tran-ris})}, \notag \\ 
	& \mathcal{L}_{k}(i, t) \leq \mathcal{L}_{max}-\mathcal{L}^{PRO}_{k}(i, t),  \forall k \in \mathcal{K} \tag{\ref{totalaten}{a}} \label{latencon}.
\end{align}

\textbf{\emph{Deriving $\mathbf{V}(i, t)$}}: Based on the ZF-based optimization algorithm for the receive beamforming matrix  \cite{21}, we can obtain $\mathbf{V}(i, t) = \boldsymbol{\mathcal{H}}_{UL}(i, t) ((\boldsymbol{\mathcal{H}}_{UL}(i, t))^{H}\boldsymbol{\mathcal{H}}_{UL}(i, t))^{-1}$,
where $\boldsymbol{\mathcal{H}}_{UL}(i, t)  = \{ \mathbf{h}_{1, a}^{UL}(i, t), ....,  \mathbf{h}_{K, a}^{UL}(i, t)\} \in \mathbb{C}^{M \times K}$.


\textbf{\emph{Deriving $\mathbf{W}(i, t)$}}: Given $\mathbf{V}(i, t)$, problem (\ref{totalaten}) can be reformulated as
\begin{align}
	&\min_{\mathbf{W}(i, t)} \sum_{k \in \mathcal{K}}{\varpi_{k}^{\iota}(i) \frac{\varsigma E_{k}(i, t)}{\upsilon_{k}^{DL}(i, t)}},  \label{downla2} \\
	s.t., ~&  \text{(\ref{tran-ris})}, \notag \\ 
	& \mathcal{L}^{DL}_{k}(i, t) \leq \mathcal{L}_{max} - \mathcal{L}^{UL}_{k}(i, t)-\mathcal{L}^{PRO}_{k}(i, t)\tag{\ref{downla2}{a}} \label{latencon2}.
\end{align}

Based on the ZF-based optimization algorithm \cite{21}, 
we can obtain as $\mathbf{W}(i, t) = \tilde{\mathbf{W}}(i, t)\mathbf{P}^{\frac{1}{2}}$,
where $\boldsymbol{\mathcal{H}}_{DL}(i, t)  = \{ \mathbf{h}_{1, a}^{DL}(i, t), ....,  \mathbf{h}_{K, a}^{DL}(i, t)\} \in \mathbb{C}^{M \times K}$, $ \tilde{\mathbf{W}}(i, t) = \boldsymbol{\mathcal{H}}_{DL}(i, t) ((\boldsymbol{\mathcal{H}}_{DL}(i, t))^{H}\boldsymbol{\mathcal{H}}_{DL}(i, t))^{-1}$, and $\mathbf{P}$ is a diagonal matrix whose $k$-th diagonal element is the receive power at the mobile user $k \in \mathcal{K}$ within time slot $t \in \mathcal{T}_i$ under DT scene $i \in \mathcal{I}$, i.e., $p^{DL}_{k}(i, t)$. Additionally, the following constraints should be satisfied:
$
	|(\mathbf{h}_{a, k}^{DL}(i, t))^H\mathbf{w}_{k}(i, t)| = \sqrt{p^{DL}_{k}(i, t)}
$
and
$
	|(\mathbf{h}_{a, k}^{DL}(i, t))^H\mathbf{w}_{m}(i, t)| = 0, \forall k \neq m.
$
Then, with these two constraints, we reformulated problem (\ref{downla2}) as:
	\begin{align}
	&\max_{\{p^{DL}_k(i, t)\}} \sum_{k \in \mathcal{K}}{ \frac{b\log_2(1+\frac{p^{DL}_{k}(i, t)}{(\sigma^{DL}_{k})^{2}})}{ \varsigma E_k(i, t)\varpi_{k}^{\iota}(i)} },  \label{downla3} \\
	s.t., ~&   \text{(\ref{tran-ris})} \notag, \\
	& \log_2(1+\frac{p^{DL}_{k}(i, t)}{(\sigma^{DL}_{k})^{2}}) \notag \\ \geq & \frac{\varsigma E_{k}(i, t) }{b(\mathcal{L}_{max} - \mathcal{L}^{UL}_{k}(i, t)-\mathcal{L}^{PRO}_{k}(i, t))},  \forall k \in \mathcal{K} \tag{\ref{downla3}{a}} \label{latencon3}.
\end{align}

We apply the water-filling algorithm to derive problem (\ref{downla3}). The closed-form solution  can be obtained as $p^{DL}_k(i, t) = \frac{1}{v_k(i, t)}\max\{\frac{1}{\varrho} - v_k(i, t) (\sigma^{DL}_{k})^{2}, v_k(i, t) p^{DL, min}_k(i, t) \}$, where $v_k(i, t)$ is the $k$-th diagonal element of $(\tilde{\mathbf{W}}(i, t))^{H}\tilde{\mathbf{W}}(i, t)$, $\varrho$ is normalization factor which is selected for ensuring that $\sum\nolimits_{k \in \mathcal{K}}\max\{\frac{1}{\varrho} - v_k (\sigma^{DL}_{k})^{2}, v_k p^{DL, min}_k \} = P^{max}$. Besides, $p^{DL, min}_k(i, t) = (\sigma^{DL}_{k})^{2} (2^{\frac{\varsigma E_{k}(i, t) }{b(\mathcal{L}_{max} - \mathcal{L}^{UL}_{k}(i, t)-\mathcal{L}^{PRO}_{k}(i, t))}} - 1)$ is the minimum received power constraint.

\subsection{Implementation of PG-ZFO}
\subsubsection{Offline Training}
We provide a dataset $\boldsymbol{\mathcal{D}}^{tra} = \{\mathcal{D}(1^{tra}), \mathcal{D}(2^{tra}), ..., \mathcal{D}(|\mathcal{I}^{tra}|)\}$ for offline training. Each $\mathcal{D}(i^{tra}) \in \boldsymbol{\mathcal{D}}^{tra}$ contains multiple episodes  corresponding to MDP $\mathcal{M}(i^{tra}) \in \boldsymbol{\mathcal{M}}^{tra}$.
We further divide  each $\mathcal{D}(i^{tra})$ into $\mathcal{D}_{\star}(i^{tra})$ for sampling the prompt, and $\mathcal{D}_{\diamond}(i^{tra})$ for sampling the most recent decision-making trajectory.

\textbf{\emph{Batching}}: For each $\mathcal{M}(i^{tra}) \in \boldsymbol{\mathcal{M}}^{tra}$, we build a minibatch $\mathcal{B}(i^{tra}) = \{In(\tau_{\star}(i^{tra}, g), \tau(i^{tra}, t_g, g))\}_{g = 1}^{G}$ includes $G$ input tokens. Each token $In(\tau_{\star}(i^{tra}, g), \tau(i^{tra}, t_g, g)), g \in \{1, ..., G\}$ consists of a prompt $\tau_{\star}(i^{tra}, g)$ sampled from $\mathcal{D}_{\star}(i^{tra})$, and a most recent decision-making trajectory $\tau(i^{tra}, t_g, g)$ sampled from $\mathcal{D}_{\diamond}(i^{tra})$. By combining all minibatches, we can obtain a batch $\mathcal{B} =\{\mathcal{B}(i^{tra}), \forall i^{tra} \in \mathcal{I}^{tra}\}$.

\textbf{\emph{Model training}}: We allow the prompt-guided decision transformer $\mathcal{X}_{\omega}$ to traverse all input tokens in the batch $\mathcal{B}$.  For each input token, the prompt-guided decision transformer generates a set of decisions. Then, we combine them as a vector of generated decisions, represented as $\boldsymbol{\vartheta}^{gen}$. After that, we compute the mean-squared error (MSE) loss $L_{MSE}$ between $\boldsymbol{\vartheta}^{gen}$ and the corresponding decisions $\boldsymbol{\vartheta}$ in $\mathcal{B}$, which is calculated as  
$L_{MSE} = \frac{1}{|\mathcal{B}|}\sum(\boldsymbol{\vartheta}^{gen} - \boldsymbol{\vartheta})^{2}.$
With the aim of minimizing $L_{MSE}$, we update the prompt-guided decision transformer using Adam optimizer with learning rate $\beta$.

\subsubsection{Online Execution}
In the online execution of PG-ZFO, we integrate the offline-trained prompt-guided decision transformer $\mathcal{X}^{*}_{\omega}$ with the ZF-based optimization algorithm.

\textbf{\emph{Input token acquisition}}: Specifically, for any $\mathcal{P}(i) \in \bm{\mathcal{P}}$, we first reformulate it as MDP $\mathcal{M}(i) \in \bm{\mathcal{M}}$. Then, we obtain the prompt $\tau_{\star}(i)$ through a deep reinforcement learning (DRL) integrated with ZF-based optimization algorithm based prompt acquisition approach \cite{21}. Then, we take current state $s(i, t)$ and the RTG $\hat{r}(i, t) = \hat{R}(i) - \sum\nolimits_{j = 1_{\star}}^{j=T_{\star}} r_{\star}(i, j)$, where current time slot $t = T_{\star} + 1$, forming the most recent decision-making trajectory $\tau(i, t)$ with sequence length of $L$. 

\textbf{\emph{Action generation}}: We input $In(\tau_{\star}(i), \tau(i, t))$ to PG-ZFO, which use the prompt-guided decision transformer $\mathcal{X}^{*}_{\omega}$ to generate $\vartheta(i, t) = \{\boldsymbol{\Theta}(i, t), E_k(i, t), f_{k}(i, t)\}_{k \in \mathcal{K}}$. Then, it uses ZF-based optimization algorithm to derive $\kappa(i, t) = \{\mathbf{V}(i, t), \mathbf{W}(i, t)\}$ by taking $\vartheta(i, t)$ as the input. Subsequently, PG-ZFO outputs the action $a(i, t) = \{\vartheta(i, t), \kappa(i, t) \}$ for the MDP $\mathcal{M}(i) \in \bm{\mathcal{M}}$, obtaining the new RTG and next state, and update the input. We interactively conduct such decision-making process until reach $T_i$.

\section{Simulation Results}

	\begin{table}[!t]
	\footnotesize
	\renewcommand{\arraystretch}{1.1}
	\centering
	\caption{Main simulation setting parameters.} \label{Set. Parameters}
	\scalebox{0.97}{
		\begin{tabular}{l|l|l|l}
			\hline
			\textbf{Parameter}        & \textbf{Value} & \textbf{Parameter}        & \textbf{Value} \\ \hline\hline
			 $K$	& 10  &  $M$	& 64 \\
			 $N$	& 16 &  $B$ & 2 MHz \\
			 $\xi$ & 8 Mb  &  $P^{max}$ & 43 dBm \\
			$p_{k}^{UL}$ & 500 mW  &  $C$ 	& 10 GHz \\
			$c$ & 50 cycle/bit &  $D_k(i)$  &  [0.1, 1.0] MB\\
			$E_{k}(i)$ &  [1.0, 2.0] &  $E_{th}$ & 1.0 \\
		 $\varsigma$  & 1 MB &  $\mathcal{L}_{max}$ & 0.5 s \\ 
		 $(\sigma^{UL}_k)^2$ & -60 dBm & $(\sigma^{DL}_k)^2$ & -50 dBm \\
		 $\rho$ & -20 dB & $\{G_{k,r}, G_{r, k}\}_{k = 1}^{K}$ & 8 dB \\
		 $G_{r, a}$ & 6 dB & $G_{a, r}$ & 7 dB \\
		 $T_i$ & 100 & $T_{\star}$ & 10 \\
		 $\mathcal{G}$ & 12 &  $L$ & 20 \\ 
		$G$	& 16  &  $\beta$	& 1e-4 \\

			\hline
		\end{tabular}
	}
\end{table}

\subsection{Experimental Settings}
We consider a RIS-assisted DT interaction system in a $100m \times 100m$ area, where the DT server is located at $\mathbf{q_a} = [0m, 0m, 40m]$ and the RIS is located at $\mathbf{q_r} = [75m, 100m, 20m]$. The DT server maintains a DT model that evolves in response to the uncertain changes of an educational system to provide DT interaction services \cite{77}.  Besides, to construct the dataset $\boldsymbol{\mathcal{D}}^{tra}$, we take $|\mathcal{I}^{tra}|=40$ DT scenes arisen from the DT model and collect 100 episodes per scene-specific problem $\mathcal{P}(i^{tra})$. Additionally, 3 distinct DT scenes from $\mathcal{I}$ are selected as examples for performance evaluation, with corresponding problems denoted as $\mathcal{P}(1^{new}), \mathcal{P}(2^{new}), \mathcal{P}(3^{new})$. 
All simulations are implemented using PyTorch and Huggingface Transformers on an NVIDIA RTX 4090 GPU. 
Table \ref{Set. Parameters} lists the values of main simulation setting parameters, and all simulation results are obtained by taking averages over 1000 runs.

The following schemes are simulated as benchmarks.

\textbf{Rigid optimization method (ROM)} \cite{88}: ROM tailors a solution policy for a random historical scene-specific problem through DRL, and then the solution policy is deployed  to solve any  new scene-specific problem.

\textbf{Decision transformer without prompt (DF-WP)} \cite{38}:
DF-WP is first offline trained on the same dataset used in PG-ZFO, and then deployed for online execution to solve any  new scene-specific problem. SDF does not consider the augmentation of prompts during either offline training and online execution.

\subsection{Performance Evaluations}\label{perform_e}
\begin{figure}[!t]
	\centering
	\includegraphics[width=2.95in]{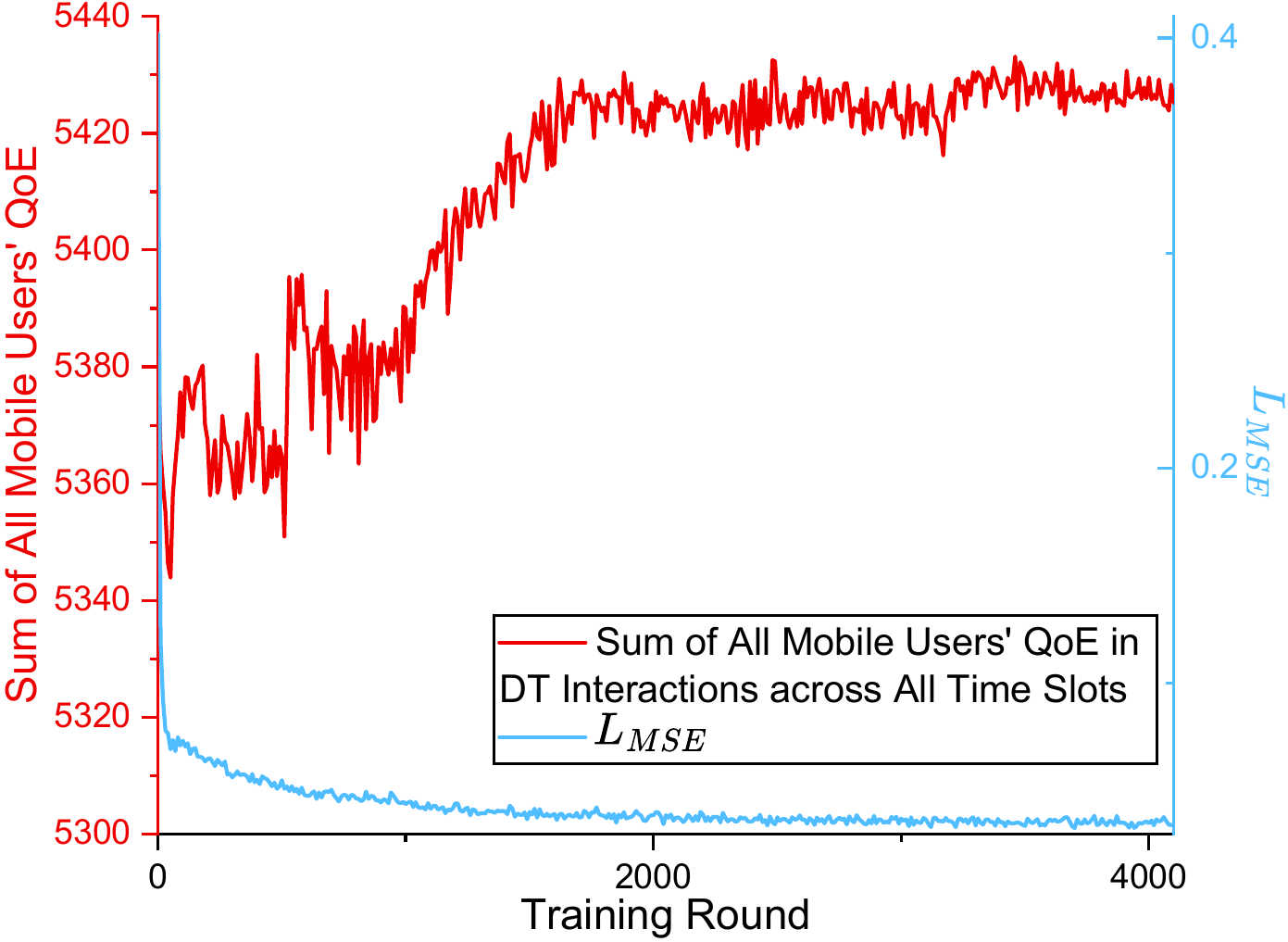} 
	\caption{Convergence of the proposed PG-ZFO approach.}\label{R-Step-QoE}
\end{figure}

\begin{figure}[!t]
	\centering
	\includegraphics[width=2.6in]{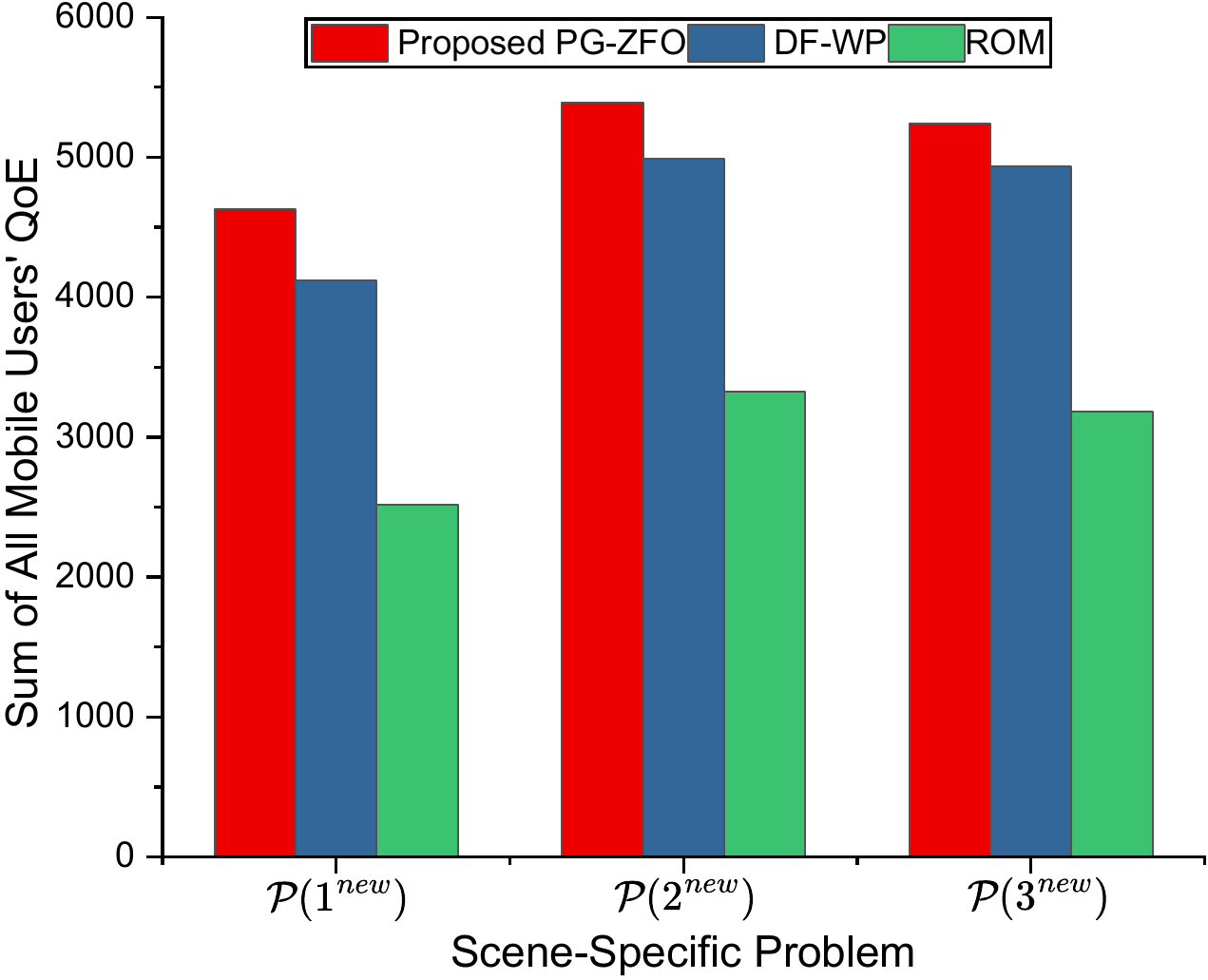}
	\caption{Illustration of the sum of all mobile users' QoE of DT interaction across all time slots under various DT scenes within $\mathcal{I}$.}\label{R-Sce-QoE}
\end{figure}

\begin{figure}[!t]
	\centering
	\includegraphics[width=2.6in]{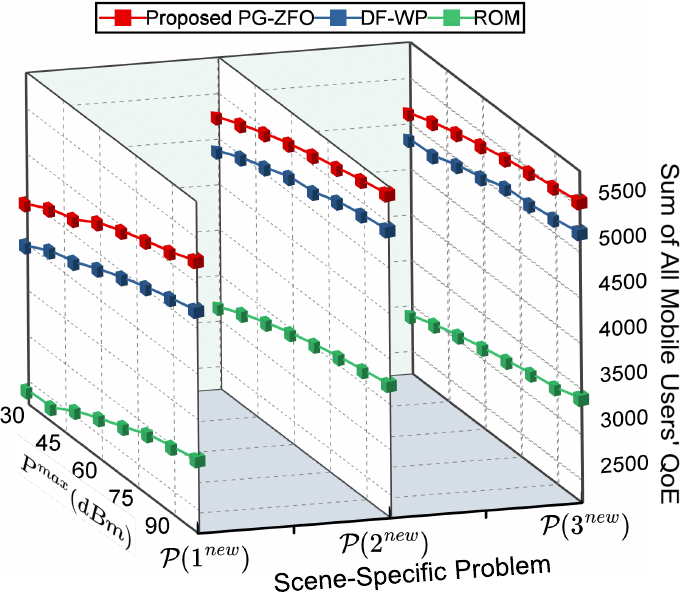}
	\caption{Comparison of the sum of all mobile users' QoE of DT interaction across all time slots w.r.t. maximum transmit power.}\label{Power-QoE}
\end{figure}


Fig. \ref{R-Step-QoE} illustrates the convergence of PG-ZFO. The MSE loss $L_{MSE}$ decreases steadily and converges to a small value. This indicates that PG-ZFO learns a generalized solution policy applicable across historical scene-specific problems $\{\mathcal{P}(i^{tra})\}_{\forall i^{tra} \in \mathcal{I}^{tra}}$. Additionally, we evaluate the PG-ZFO in the unseen problem $\mathcal{P}(i+1)$ during the offline training. The results shows that  $\sum_{t\in\mathcal{T}_{i+1}}\sum_{k\in\mathcal{K}}QoE(\mathcal{W}_k(i+1), t)$, increases as $L_{MSE}$ decreases and stabilizes when $L_{MSE}$  converges. This indicates that PG-ZFO generalizes well to unseen problems, attributed to its use of prompts in PG-ZFO, which enriches the representation of scene-specific information and enhance generalization.

Fig. \ref{R-Sce-QoE} illustrates the  sum of all mobile users' QoE in DT interactions across all time slots under various DT scenes in $\mathcal{I}$, optimized by ROM, DF-WP and the proposed PG-ZFO.
The proposed PG-ZFO outperforms DF-WP, as its prompts provide rich scene-specific information to guide action generation, improving policy generalization. DF-WP performs better than ROM, as its 
RTG mechanisms partially captures scene-specific information, helping it maximize rewards, which helps guide it to generate actions toward the maximum rewards $\hat{\mathcal{R}}(i)$ that expected to be achieved in the $\mathcal{M}(i) \in \bm{\mathcal{M}}$.

Fig. \ref{Power-QoE} examines the sum of all mobile users' QoE in DT interactions across all time slots under various DT scenes within $\mathcal{I}$ with different maximum transmit power $P^{max}$. 
As $P^{max}$ increases, 
 $\sum_{t\in\mathcal{T}_i}\sum_{k\in\mathcal{K}}QoE(\mathcal{W}_k(i), t), i \in \mathcal{I}$ increases because higher $P^{max}$ reduces downlink latency $\mathcal{L}^{DL}_{k}(i, t)$, enhancing mobile users' objective experience. The proposed PG-ZFO outperforms both ROM and DF-WP. The explanations for these results are similar to those for Fig. \ref{R-Sce-QoE}.

\section{Conclusion}
In this paper, we investigate a QoE-aware resource allocation problem for RIS-assisted DT interaction with uncertain evolution. 
We aim to maximize the sum of all mobile users' QoE in DT interactions across all scene-specific problems, by jointly determining phase shift matrix, receive/transmit beamforming matrix, rendering resolution configuration and computing resource allocation. To solve this complicated problem, 
 we propose a novel GAI-aided approach, called PG-ZFO. Simulation results show that, the proposed PG-ZFO has superior performance compared to counterparts.

\section{Acknowledgments}
This work was supported by Postgraduate Research \& Practice Innovation Program of Jiangsu Province No.KYCX24\_0596.

\bibliographystyle{IEEEtran}

\bibliography{ref}

\end{document}